# ASTROD I Charging Simulation and Disturbances

GANG BAO[1,2], D N A SHAUL[3], H M ARAUJO[3], WEI-TOU NI[1], T J SUMNER[3] AND LEI LIU[1,2]

1 Purple Mountain Observatory, Chinese Academy of Science, Nanjing, China, 210008

2 Graduate University of the Chinese Academy of Science, Beijing, China, 100049

3 Department of Physics, Imperial College London, London, SW7 2BZ, UK

Email:bgastro@pmo.ac.cn

ASTROD I is planned as a single spacecraft mission. It will use interferometric and pulse ranging techniques between the spacecraft and ground stations, to make high precision measurements of the parameters that describe the solar system, and to test relativistic gravity with improved accuracy. At the heart of the spacecraft is a test mass, which the spacecraft will follow using a drag-free control system. The mission critically depends on maintaining the geodesic motion of the test mass. Charging of the test mass due to cosmic rays and solar particles will disturb its geodesic motion. We have modelled the charging process using the GEANT4 toolkit and a simplified, geometrical model and estimate that the ASTROD I test mass will charge positively, at a rate of 24 ± 7 $e^+$/s, due to cosmic ray protons and alpha particles ($^3$He and $^4$He) at solar minimum. We have used the results of this simulation to estimate the magnitude of disturbances associated with test mass charging, for the worst-case scenario, taking into account uncertainties in the model and potential charging contributions from minor cosmic-ray components.

## 1. INTRODUCTION

The ASTROD I mission concept is based around a single, drag-free spacecraft and laser interferometric ranging and pulse ranging with ground stations. It is the first step towards realising the ASTROD mission (the Astrodynamical Space Test of Relativity using Optical Devices) [1-3]. The scientific goals of ASTROD I include measuring relativistic parameters with better accuracy, improving the sensitivity achieved in using the optical Doppler tracking method for detecting gravitational waves, and measuring many solar system parameters more precisely. The spacecraft will be 3-axis stabilized, with a 3-axis drag-free test mass in the center. The test mass will be surrounded by electrodes on all six sides, to capacitively sense its motion relative to the spacecraft. Micro-thrusters on the spacecraft would then be used to force it to follow the test mass. The spacecraft will be launched into a low earth orbit and from there will be injected, using a medium-size ion thruster, directly into an orbit around the Sun. This solar orbit will initially have a period of 282 days, but, after two encounters with Venus, the period will be shortened to about 165 days. After about 370 days after launch, the spacecraft will reach the far side of the Sun. At this distance from the Earth, it will be possible to determine relativistic parameters with unprecedented accuracy by measuring the time of arrival of laser pulses transmitted between ground stations and the spacecraft with 10ps accuracy. If the residual acceleration noise can be reduced to the levels shown in Fig. 1, orbit simulations indicate that 400 days after launch, in a period of 450 days, both the light deflection/retardation parameter, γ, and the nonlinear relativistic-gravity parameter, β, can be determined to $10^{-7}$, which represents 200 and 1000 fold improvements, respectively. Further, in this period, the solar quadrupole parameter, $J_2$, can be determined to $10^{-8}$, and 1-3 orders of magnitude improvements in solar system parameters, such as the masses of planets, can be made.

---

[1] Purple Mountain Observatory, Chinese Academy of Science, Nanjing, China, 210008, Email:bgastro@pmo.ac.cn
[2] Graduate University of the Chinese Academy of Science, Beijing, China, 100049
[3] Department of Physics, Imperial College London, London, SW7 2BZ, UK

The basic payload configuration of ASTROD I is a cylindrical spacecraft, of diameter 2.5 m and height 2 m, whose surface is covered with solar panels [1-4]. In orbit, the cylindrical axis would be perpendicular to the orbital plane, with a telescope pointing towards a ground laser station. The effective area for receiving sunlight would be about 5 m$^2$, which can generate over 500 W of power. The total mass of the spacecraft is 300-350 kg and that of payload is 100-120 kg. The test mass is a 1.75 kg, rectangular parallelepiped, made of an extremely low magnetic susceptibility ($<10^{-5}$) Au-Pt alloy, to minimize magnetic disturbances [1, 5]. The ASTROD I residual acceleration noise target is

$$S_{\Delta a}^{1/2}(f) = 3 \times 10^{-14} \left[ \frac{0.3 \text{ mHz}}{f} + 30(\frac{f}{3 \text{ mHz}})^2 \right] \text{ ms}^{-2}\text{Hz}^{-1/2}, \quad (1)$$

over the frequency range of 0.1 mHz $< f <$ 100 mHz. This is compared to the LISA Pathfinder LISA Technology Package (LTP) [6] and LISA [7] noise curves in Fig. 1.

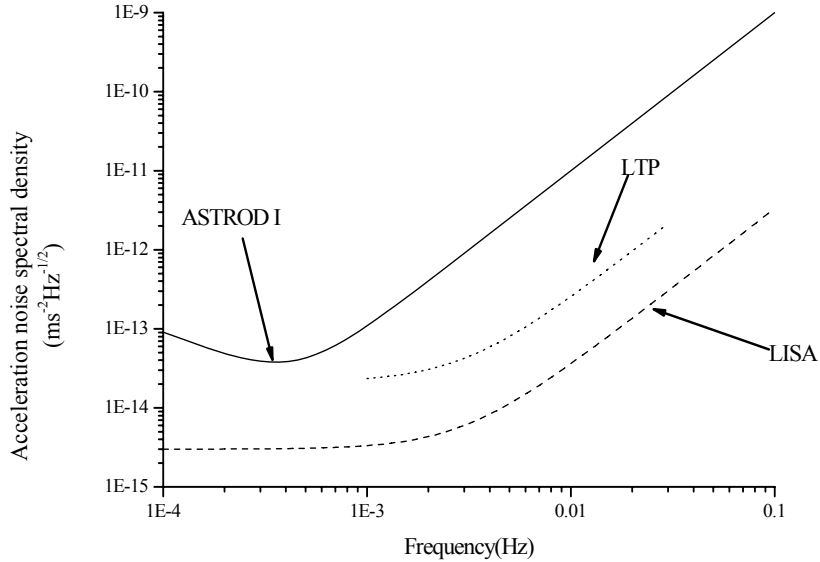

Figure 1. A comparison of the target acceleration noise curves of ASTROD I, the LTP and LISA.

High-energy, cosmic rays and solar energetic particles (SEPs) easily penetrate the light structure of spacecraft transferring heat, momentum and electrical charge to the test mass [8]. Electrical charging is the most significant of these disturbances. Any charge on a test mass will interact with the surrounding conducting surfaces through Coulomb forces. Further, motion of the charged test mass through magnetic fields will give rise to Lorentz forces. To limit the acceleration noise associated with these forces and meet the residual noise requirement, the test mass must be discharged in orbit.

We present here preliminary estimates of the ASTROD I net test mass charging rate and shot noise due to cosmic rays, based on simulations using the GEANT4 toolkit [9]. The derived test mass charging characteristics depend on the geometry, materials, incident flux and physics processes that are used in the model. These are described in section 2. The simulation results are described in section 3. In section 4, these results are used to estimate the magnitude of the Coulomb and Lorentz disturbances due to test mass charging for ASTROD I.

## 2. CHARGING SIMULATION

### 2.1. Geometry and Materials

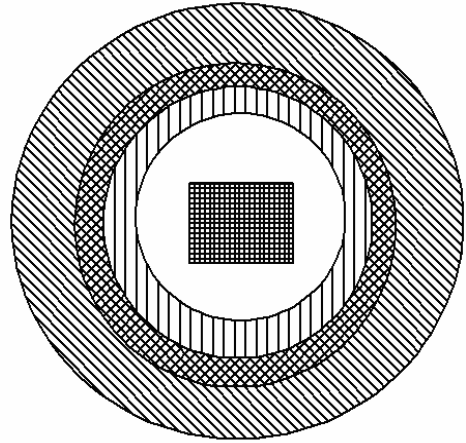

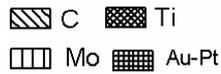

Figure 2. ASTROD I geometry model.

The geometry model used in this study is sketched in Fig. 2. The test mass is a 50 ×50×35 mm$^3$ Au-Pt alloy (density: 20 g/cm$^3$) rectangular parallelepiped at the center of the model, surrounded by 3 concentric, spherical shells. The material of the innermost shell is molybdenum (used to simulate the electrodes); that of the middle shell is titanium (used to simulate the sensor enclosure); that of the outermost shell is carbon (used to simulate the structure of spacecraft, equipment, battery etc.) The unshaded region between the test mass and molybdenum shell in Fig. 2 is a vacuum ($1.0 \times 10^{-25}$ g/cm$^3$). The thicknesses and densities of the 3 material shells used in the model are reported in Table 1. It should be noted that the mass of this model is about 6 kg, whereas the total mass of the ASTROD I spacecraft is 300-350 kg. The discrepancy in mass is due to the fact that the actual size of the spacecraft is much larger than this model. This is not expected to affect the derived net charging rate and the charging noise substantially, based on a trial run in which a 261kg, 10mm thick, 2 m diameter, carbon, concentric shell was added to the model.

| Material | Density (g/cm$^3$) | Radius (mm) | Thickness (mm) |
|---|---|---|---|
| Molybdenum | 10.28 | 46 | 6 |
| Titanium | 4.54 | 51 | 5 |
| Carbon | 2.10 | 71 | 20 |

Table 1. Characteristics of each material layer (the radius in table 1 is the distance from the center of the test mass to the outer surface of each layer)

## 2.2. Incident Flux

The ASTROD I orbit varies between 0.4864 AU and 1.0145 AU, in radial distance from the Sun, and between + 0.837 degrees to - 0.668 degrees, in latitude from the ecliptic plane, as shown in Fig. 3 for 4 August, 2010 launch.

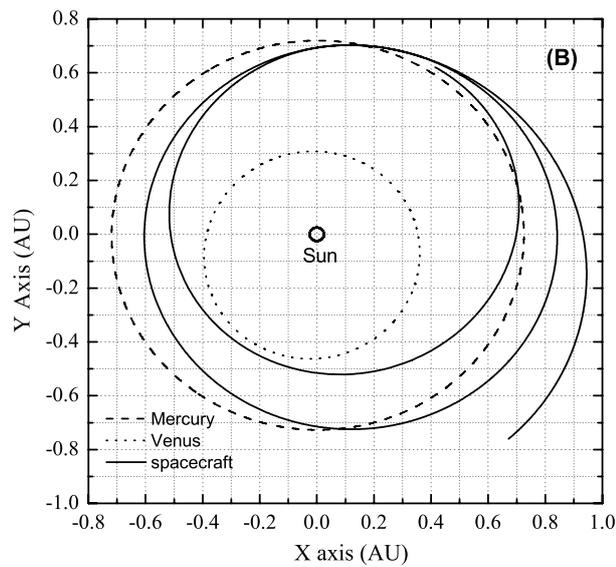

Figure 3. The ASTROD I orbit [10].

The Ulysses study [11], observed a (0.33+/-0.04)%/degree increase in cosmic ray flux towards the polar regions in both hemispheres, at solar minimum. At solar maximum, no latitudinal dependence of cosmic ray flux was found. This implies that for this simulation, the cosmic ray flux near the Earth can be adopted, without need for corrections due to the ASTROD I variations in latitude [12].

The modulation parameter at solar minimum is expected to vary between 100-200 MeV(/n) and 200-300 MeV(/n) at radial distances of 1AU and 0.5 AU, respectively. This implies just a few percent reduction in cosmic ray flux over the whole spectrum, comparing 0.5 AU to 1AU. However, at 100 MeV(/n), which is the charging threshold for ASTROD I (see section 3), the flux difference may be as large as a factor of 2 [12]. For this study, cosmic ray spectra at 1AU, at solar minimum, [13] have been used, as this is expected to give an estimate of the upper limit to charging disturbances over the entire ASTROD I orbit, in solar quiet conditions.

The particles simulated in this paper are cosmic ray protons and alpha particles ($^3$He and $^4$He), which represent approximately 98% of the total amount of cosmic rays. The effect of cosmic ray fluxes of other particle species, such as C, N, O and $e^-$, are discussed in section 3. The differential energy spectra for the proton and helium fluxes at solar minimum [13] that have been used in this simulation are plotted in Fig. 4 and Fig. 5.

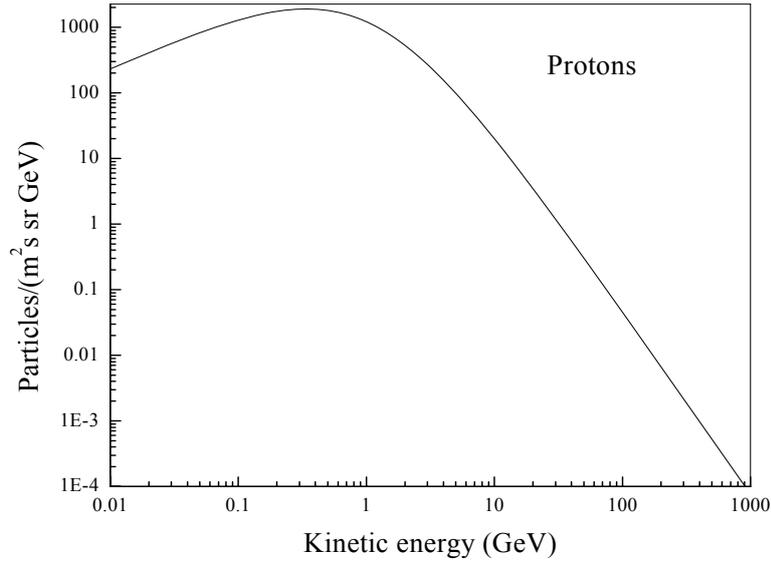

Figure 4. Differential energy spectra for cosmic ray protons at solar minimum [13].

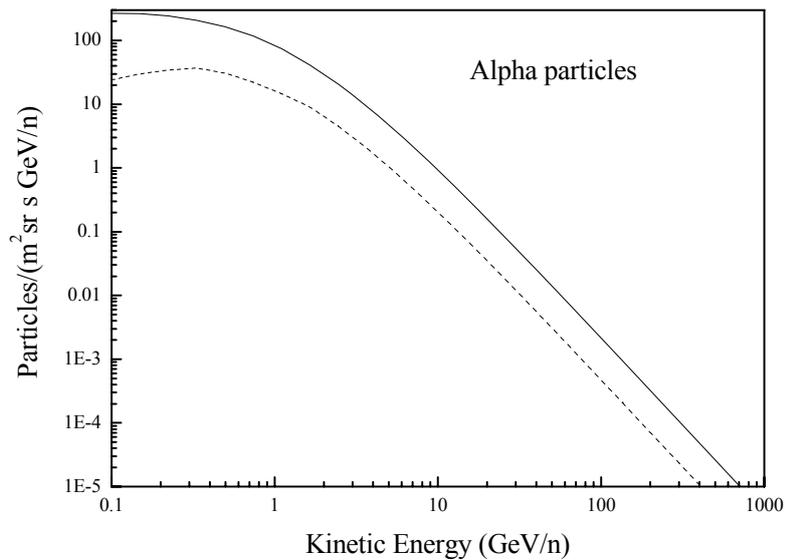

Figure 5. Differential energy spectra for cosmic ray alpha particles at solar minimum. In the figure, the dotted line is for $^3$He and the solid line is for $^4$He [13].

## 2.3. Physics Modeling

GEANT4 is a toolkit for the simulation of the passage of particles through matter. In this simulation, cosmic rays are emitted isotropically, irradiating the entire spacecraft uniformly, from an outer spherical shell (radius 99 mm). The primary energies are sampled in the range of 0.01-1000 GeV for protons and 0.1-1000 GeV/n for alpha particles, which are from the distributions plotted in Fig. 4 and Fig. 5. The physics processes used in our simulation are broadly divided into electromagnetic (EM) and hadronic interactions. As a result of their high energy and hadronic nature, cosmic ray interactions bring forth complex nuclear reactions which have large final-state multiplicities, producing a plethora of secondaries. A low energy threshold of 250 eV was adopted for secondary particle production in our simulation. Fluorescence and non-radiative (Auger) atomic deexcitation have been implemented. The hadronic physics is mainly implemented by elastic and inelastic scattering processes. The inelastic reactions were based on the LEP and HEP parameterized models. The inelastic reactions also use evaporation models to treat the deexcitation of nuclei with A>16, comprising gamma emission, fragment evaporation ($p, n, \alpha, {}^2H$ and ${}^3H$) and fission of heavier residual nuclei. A variety of decay, capture and annihilation processes has also been included in our physics processes list. The charging potential of several additional physics processes, such as the kinetic emission of very low energy electrons, has not been modeled in the present simulation, but has been estimated based on LISA studies [14].

## 3. SIMULATION RESULTS

We have run three independent GEANT4 simulations to determine the charging of the ASTROD I test mass by cosmic ray protons, ${}^3$He and ${}^4$He, respectively. In total, about 280,000 events were simulated. The details of each event that resulted in test mass charging were recorded, including the event time, net charge deposited on the test mass and the energy of the primary. The variation of the net test mass charge with time, due to the proton, ${}^3$He and ${}^4$He fluxes are shown in Fig. 6, Fig. 7 and Fig. 8, respectively. The straight lines in these figures correspond to least squares fits of this data, giving mean net charging rates attributable to the proton, ${}^3$He and ${}^4$He fluxes of 19.2 $\pm$ 0.5 $e^+$/s, 0.69 $\pm$ 0.05 $e^+$/s and 4.3 $\pm$ 0.2 $e^+$/s, respectively. The errors quoted account for only the Monte

Carlo (MC) uncertainty, calculated by combining the Poisson variances for the occurrence of each net event charge. This indicates that ~97% of the charge accumulated comes from primary cosmic ray protons and $^4$He, and all three fluxes lead to positive charging of the test mass. Although the proton flux dominates these rates, $^4$He, which constitutes only 8% of the total cosmic rays flux, is responsible for ~18% of the test mass charging. A histogram of the net charge deposited in an event is given in Fig. 9, for the proton data set, showing that most events result in the transfer of one unit of charge. The effects of the positive and negative charging currents cancel to some extent, but an imbalance in these currents gives rise to the net positive charging rate.

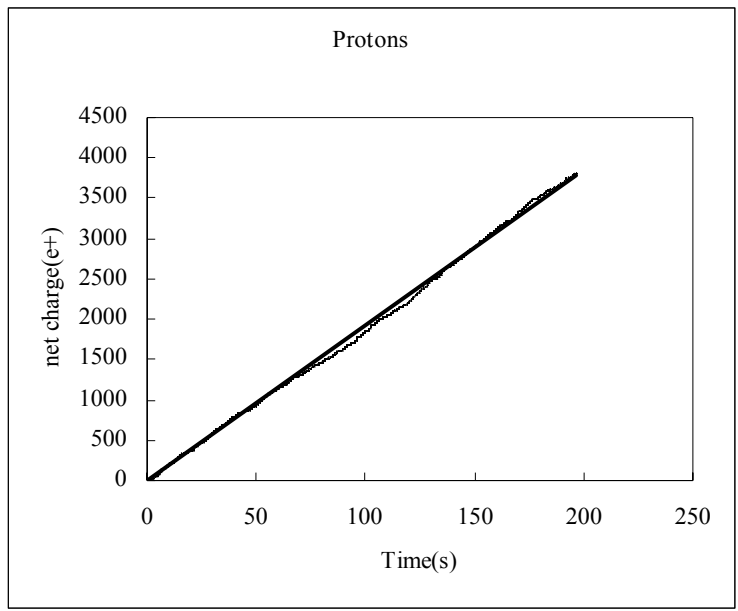

Figure 6. The charging timeline for protons. The straight line corresponds to a least squares fit to the data.

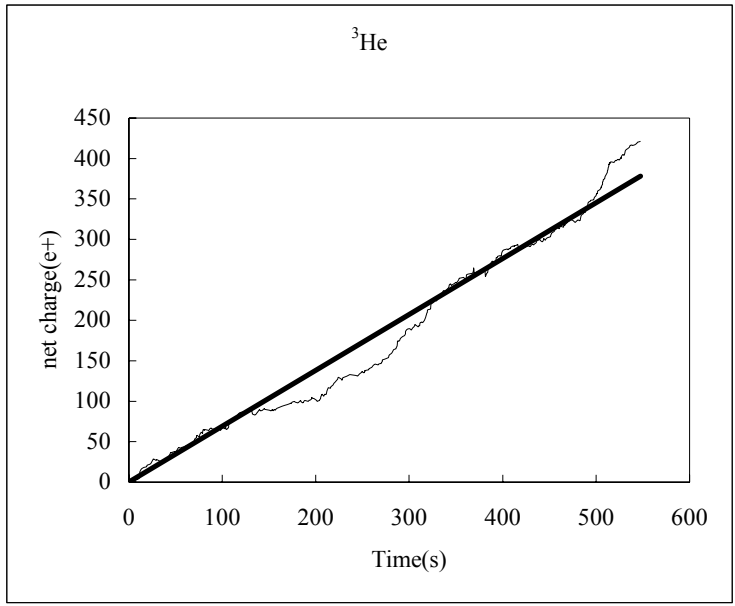

Figure 7. The charging timeline for $^3$He. The straight line corresponds to a least squares fit to the data.

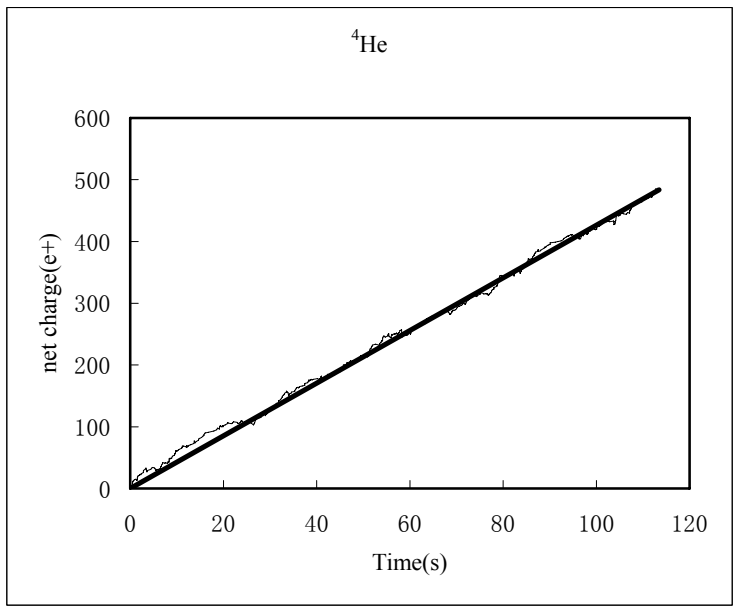

Figure 8. The charging timeline for $^4$He. The straight line corresponds to a least squares fit to the data.

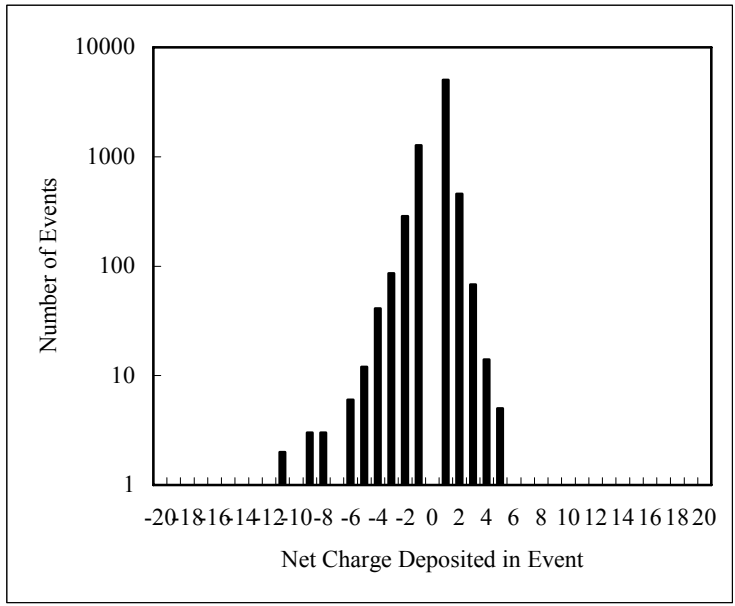

Figure 9. Histogram of the net charge deposited in an event, for incident protons. The total number of proton events simulated was 260,000.

The charging rate is plotted as a function of primary energy in Fig. 10. The low energy cut-off is due to the shielding provided by the spacecraft, which prevents incident protons with energies below ~100 MeV from charging the test mass. The most significant charging mechanism is primary cosmic ray particles stopping in the test mass. This occurs mainly for protons of energy between ~ 100 – 610 MeV. Protons with energies in excess of ~ 610 MeV have sufficient energy to traverse the distance through the spacecraft to the test mass and the longest path through the test mass, without being stopped. This explains the peak observed in this energy interval in Fig. 10.

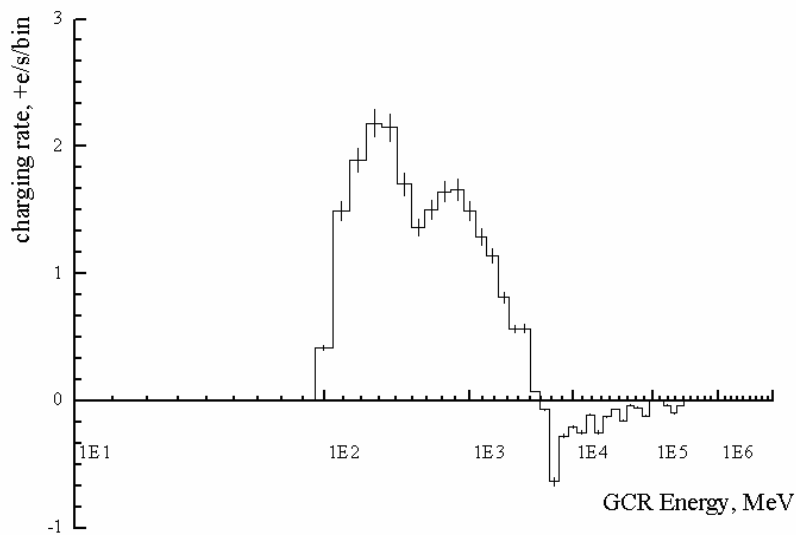

Figure 10. Charging rate of test mass as a function of primary energy.

In addition to the MC uncertainty, an error of ± 30% in the net charging rate should be considered, to account for uncertainties in the GCR spectra, physics models and geometry implementation. Further, a potential contribution to the charging rate of 28.4 $e^+$/s from kinetic low energy secondary electron emission should be considered, based on LISA studies [14].

Following [8], the charging flux is considered to be made up of independent currents, $I_q$, each composed solely of charges $eq$, with shot noise of single-sided spectral density $S_q = \sqrt{2qeI_q}$, where $e$ is the magnitude of electron charge. The total noise, $S_R$, is then given by the quadrature sum of $S_q$, over all values of $q$. At solar minimum, considering the MC currents alone, $S_q$=13.2 e/s Hz $^{-1/2}$. Low energy secondary electron emission is estimated to contribute an extra 10.3 e/s Hz $^{-1/2}$, based on the LISA study [14]. Integrating in the time domain gives the charging fluctuations at frequency $f$,

$$S_Q(f) = S_R / 2\pi f, \qquad (2)$$

Based on the LISA study [15], the effect of cosmic ray fluxes of particle species not included in this simulation is expected to increase the net charging rate by ~4.2% and the charging noise by ~11%. The charging rate and noise contributions from the different sources mentioned above are summarized in Table 2. Using these figures, the worst case charging rate and noise are estimated to be 62 $e^+$/s and 18 e/s Hz $^{-1/2}$, respectively.

| Particle | Proton | $^3$He | $^4$He | Secondary electron | Other species (C, N, O, $e^-$) | Uncertainty (MC+30%) |
|---|---|---|---|---|---|---|
| Charging rate ($e^+$/s) | 19.20 | 0.69 | 4.30 | 28.36 | 1.02 | 7.28 |
| Charging noise (e/s /Hz$^{0.5}$) | 11.52 | 2.41 | 6.04 | 10.30 | 6.43 | |

Table 2. The charging rate and noise contributions from the different sources

# 4. CHARGING DISTURBANCES

The accumulation of charge on the test mass will give rise to acceleration noise and coherent Fourier components in the measurement bandwidth through both Coulomb and Lorentz interactions. Further, the position dependence of Coulomb forces can modify the effective stiffness, or coupling between the test mass and the spacecraft. These disturbances are evaluated in this section.

## 4.1. Coulomb Noise and Stiffness

The charge-dependent Coulomb acceleration $a_{Qk}$ in direction $\hat{k}$ is given by

$$a_{Qk} = \frac{Q^2}{2mC_T^2}\frac{\partial C_T}{\partial k} + \frac{QV_T}{mC_T}\frac{\partial C_T}{\partial k} - \frac{Q}{mC_T}\sum_{i=1}^{N-1}V_i\frac{\partial C_{i,N}}{\partial k}, \quad (3)$$

The first two terms in equation (3) are dependent on the overall sensor geometric symmetry, through $\frac{\partial C_T}{\partial k}$, and the third term is dependent on the symmetry of the sensor voltage distribution. The corresponding acceleration noise, $\delta a_{Qk}$, due to random fluctuations of the test mass position relative to the spacecraft, $\delta k$, of the potentials of the conductors that surround the test mass, $\delta V_i$, and of the test mass free charge $\delta Q$, is given by

$$\delta a_{Qk}^2 = (\frac{\partial a_{Qk}}{\partial k})^2 \delta k^2 + \sum_{i=1}^{N-1}(\frac{\partial a_{Qk}}{\partial V_i})^2 \delta V_i^2 + (\frac{\partial a_{Qk}}{\partial Q})^2 \delta Q^2, \quad (4)$$

where $k$ is a displacement in direction $\hat{k}$; $m$ is the mass of the test mass; $Q$ is the free charge accumulated on the test mass; $C_{i,j}$ is the capacitance between conductors $i$ and $j$ which surround the test mass; $V_i$ is the potential to which conductor $i$ is raised; $C_T \equiv \sum_{i=1}^{N-1}C_{i,N}$ is the coefficient of capacitance of the test mass, which is defined as the $N^{th}$ conductor, with potential $V_N = \frac{Q}{C_T} + V_T$, and $V_T \equiv \frac{1}{C_T}\sum_{i=1}^{N-1}C_{i,N}V_i$ [16]. The estimates for acceleration noise have assumed typical parameter values for the ASTROD I mission: $Q$ was taken as the amount of charge accumulated in 1 day, assuming a net test mass charging rate of 62 e$^+$/s, which corresponds to the MC rate, with error margins, estimated contributions from particle species not included in the MC model and the potential contribution from kinetic low energy secondary electron emission, that is likely to almost cancel in the actual sensor [14], added; $m = 1.75$ kg; mean voltages on opposing conductors $V_i = 0.5$ V; the potential difference between conductors on opposing faces of the sensor compensated to 10 mV; the asymmetry in gap across opposite sides of test mass $= 10$ $\mu$ m; capacitances and capacitance gradients were calculated using parallel plate approximations: $c_T = 53$ pF; $V_T = 0.5$ V; position noise $\delta k = 1 \times 10^{-7}$ mHz$^{-1/2}$; voltage noise $\delta V_i = 1 \times 10^{-4}$ VHz$^{-1/2}$ and charge noise $\delta Q = 4.6 \times 10^{-15}$ CHz$^{-1/2}$, which includes, as for the charging rate, the unmodelled contributions. The magnitude of the acceleration noise associated with charge, due to random fluctuations of the test mass position relative to the spacecraft and of the potentials of the conductors that surround the test mass are each estimated to be $\sim 5 \times 10^{-15}$ ms$^{-2}$Hz$^{-1/2}$. The acceleration noise associated with charging shot noise increases with decreasing frequency (see equation 2), and at 0.1 mHz, this is estimated to be $\sim 4 \times 10^{-15}$ ms$^{-2}$Hz$^{-1/2}$. The total Coulomb acceleration noise due to the test mass charging is $\sim 8 \times 10^{-15}$ ms$^{-2}$Hz$^{-1/2}$. The total noise estimated here is a factor of $\sim 10$ less than the ASTROD I acceleration noise target. The stiffness associated with test mass charging, $S_{QK}$, is given by $S_{QK} = -m \frac{\partial a_{QK}}{\partial k}$. Using the parameter values listed above gives $S_{QK} \sim -7 \times 10^{-8}$ s$^{-2}$.

## 4.2. Lorentz Noise

Lorentz effects arise from the motion of the test mass through the interplanetary magnetic field, $\vec{B}_I$, and its residual motion through the field generated within the spacecraft, $\vec{B}_S$. The test mass will be housed in a conducting enclosure, which will reduce the effect of the interplanetary field, via the Hall effect, with efficiency $\eta$. Hence, to first order, the Lorentz acceleration noise, $a_L$, is given by

$$m^2(a_L)^2 \approx (\eta Q V_I \delta B_I)^2 + (\eta Q \delta V_I B_I)^2 + (Q \delta V_S B_S)^2 + (\eta \delta Q V_I B_I)^2. \quad (5)$$

where $V_I$ is the speed of the test mass through the interplanetary field; $\delta V_I$ and $\delta V_S$ are the magnitudes of random fluctuations in the test mass velocity through the interplanetary field and relative to the spacecraft, respectively and $\delta B_I$ gives the magnitude of fluctuations in the interplanetary field [16]. $a_L$ also increases with decreasing frequency, and is estimated to be $2 \times 10^{-15}$ ms$^{-2}$Hz$^{-1/2}$ at 0.1 mHz, which is a factor of ~ 40 below the ASTROD I acceleration noise target. We have assumed that $Q = 62$ e$^+$/s, as in section 4.1; $\eta = 0.1$; $\bar{V}_I = 4 \times 10^4$ m/s; $\delta V_I = 4.78 \times 10^{-12}$ ms$^{-1}$Hz$^{-1/2}$; $\delta V_S = 6.28 \times 10^{-11}$ ms$^{-1}$Hz$^{-1/2}$; $\bar{B}_S = 9.6 \times 10^{-6}$ T; $|\delta B_S| = 1 \times 10^{-7}$ THz$^{-1/2}$; $\bar{B}_I = 1.2 \times 10^{-7}$ T (this is a conservative estimate of the field at 0.5AU, used to give the worst-case noise, for the ASTROD orbit) and $|\delta B_I| = 1.2 \times 10^{-6}$ THz$^{-1/2}$.

### 4.3. Coherent Fourier Components

The steady build up of charge on the test mass will give rise to coherent Fourier components in the ASTROD I frequency bandwidth [17]. Substituting $Q(t) \approx \bar{Q} t$, where $t$ is the time for which the test mass has been allowed to charge and $\bar{Q}$ is the mean charging rate, into the expressions for the Coulomb and the Lorentz accelerations gives the terms [17]:

$$f_k(t) \equiv \Xi_k t^2 \equiv \frac{\bar{Q}^2}{2mC_T^2} \frac{\partial C_T}{\partial k} t^2, \qquad (6)$$

$$e_k(t) \equiv \Theta_k t \equiv -\frac{\partial V_T}{\partial k} \frac{\bar{Q}}{m} t, \qquad (7)$$

and $\qquad l_k(t) \equiv \Phi_k t \equiv \frac{\eta \bar{Q} t}{m}(\vec{V}_I \times \vec{B}_I) \cdot \hat{k}. \qquad (8)$

Implementing the parameter values given in section 4.1 and 4.2, taking the mean charging rate as constant and assuming that the test mass is discharged once every 24 hours (as described in [17]), the spectral densities of $f_k(t)$ and $e_k(t)$ are estimated to exceed the ASTROD I acceleration noise limit over ~1% of the bandwidth (see Fig. 11). Several schemes could be used to minimize a potential loss of the ASTROD I science data, including continuously discharging the test mass, minimizing sensor voltage and geometrical offsets and through spectral analysis [17]. Variations in, for example, the mean charging rate, could result in these signals exceeding the noise target in a larger fraction of the bandwidth.

Hence, variations in these signals need to be studied carefully as they will influence the accuracy with which the solar-system and relativistic parameters can be determined.

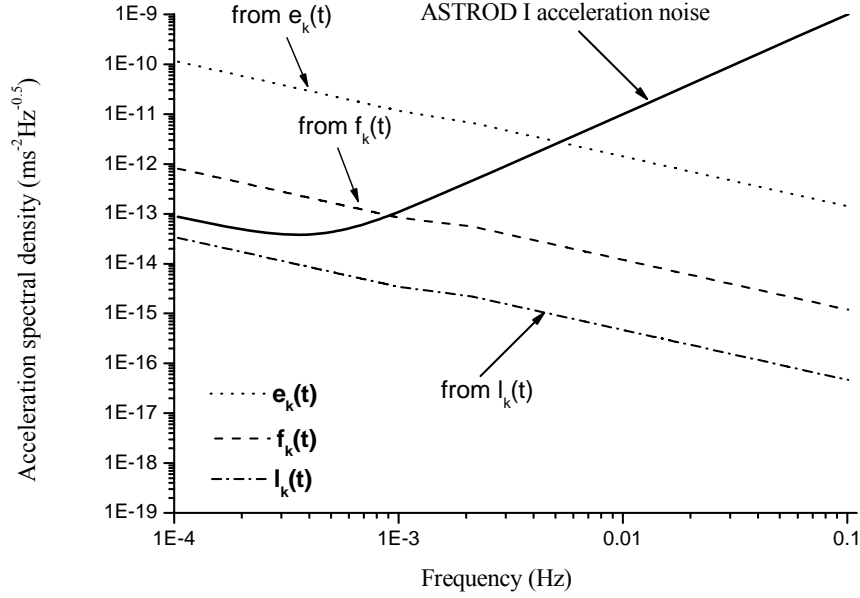

Figure 11. The curves trace the spectral densities at the primary peaks of the sinc functions [17], of the coherent Fourier components, for $\tau$ =1 year: $e_k(t)$ is given by the dashed line, $f_k(t)$ is given by the dashed-dotted line and $l_k(t)$ is given by the dotted line. The bold full line gives the ASTROD I acceleration noise limit.

## 5. Conclusion

The charging of the ASTROD I test mass by cosmic ray protons and alpha particles ($^3$He and $^4$He) has been simulated using the GEANT4 toolkit. The MC model predicted a net charging rate of 24 $\pm$ 7 e$^+$/s. Although the proton flux is the dominant charging flux, $^4$He, which constitutes only 8% of the total cosmic ray flux is responsible for ~18% of this rate. We have also considered an additional net charging rate contribution of 1.02 e$^+$/s, due to particle species that were not included in the MC model, and a potential additional ~28 e$^+$/s, due to kinetic low energy secondary electron emission. There is an additional uncertainty of $\pm$ 30% in the MC net charging rate, due to uncertainties in the cosmic ray spectra, physics models and geometry implementation.

The ASTROD I acceleration noise limit target is $10^{-13}$ ms$^{-2}$ Hz$^{-1/2}$ at 0.1 mHz, which is less stringent than the LISA requirement. The magnitudes of the Coulomb and Lorentz acceleration noise associated with test mass charging increase with decreasing frequency. At the lowest frequency in the ASTROD I bandwidth, 0.1mHz, they are estimated to be $8\times 10^{-15}$ ms$^{-2}$Hz$^{-1/2}$ and $2\times 10^{-15}$ ms$^{-2}$Hz$^{-1/2}$, respectively, both well below the acceleration noise target. However, variations in the test mass charging rate will alter the spectral description of the coherent Fourier components. Hence, further work is needed to ensure that these do not compromise the quality of the science data of the ASTROD I mission.

The next stage in this work is to implement a more realistic ASTROD I geometry model and a more complete list of physics processes in the GEANT4 simulation. Future work will also include the detailed study of other potential charging mechanisms, such as x-ray processes and photon emission.

The effect of cosmic ray fluxes of particle species not included in the MC simulation needs to be verified for the ASTROD I geometry. Further, we will evaluate the variation in the ASTROD I test mass charging rate over the solar cycle, including a detailed study of SEP events, and its variation due to modulation of cosmic ray flux over the ASTROD I orbit.

**Acknowledgements**

We thank Catia Grimani (University of Urbino and INFN Florence) for providing estimates of the cosmic ray spectral intensity. During his period of study in Imperial College for this work, one of the authors (Bao Gang) was supported by the ASTROD scholarship for visiting Europe. This work is funded by the National Natural Science Foundation (Grant Nos 10343001 and 10475114) and the Foundation of Minor Planets of Purple Mountain Observatory.